\documentclass[aps, pra, showpacs, reprint]{revtex4-1}
\usepackage{graphicx}
\usepackage{amsmath}
\usepackage{amssymb}
\usepackage{xspace}

\begin{document}
\title{An atom optics approach to studying lattice transport phenomena}
\author{Bryce Gadway}
\email{bgadway@illinois.edu}
\affiliation{Department of Physics, University of Illinois at Urbana-Champaign, Urbana, IL 61801-3080, USA}
\date{\today}
\begin{abstract}
We present a simple experimental scheme, based on standard atom optics techniques, to design highly versatile model systems for the study of single particle quantum transport phenomena. The scheme is based on a discrete set of free-particle momentum states that are coupled via momentum-changing two-photon Bragg transitions, driven by pairs of interfering laser beams. In the effective lattice models that are accessible, this scheme allows for single-site detection, as well as site-resolved and dynamical control over all system parameters. We discuss two possible implementations, based on state-preserving Bragg transitions and on state-changing Raman transitions, which respectively allow for the study of nearly arbitrary single particle Abelian U(1) and non-Abelian U(2) lattice models.
\end{abstract}

\maketitle

\section{I. INTRODUCTION}

Over the past few decades, atomic, molecular, and optical (AMO) systems have played an increasingly important role in shaping our understanding of complex quantum phenomena. Precise knowledge of the microscopic properties of AMO systems, combined with unprecedented levels of control and novel diagnostic tools, have stimulated the development of several platforms - based on cold atoms~\cite{AtomsRev-NatPhys-2012}, trapped ions~\cite{IonsRev-NatPhys-2012}, and photons~\cite{PhotRev-NatPhys-2012} - for the quantum simulation of myriad physical phenomena, especially those related to condensed matter~\cite{Bloch-RMP08,Lewenstein-Mimicking-}.

For the study of single electron transport phenomena, photonic~\cite{Wiersma-LightLocal-1997,SzameitReview-2010,Quasi1D-2D-Kraus-2012A,Segev-LightLocal-2013,Rech-TopFloquet-2013,Hafezi-ImagingTop-2013} and cold atom~\cite{Moore-Qdeltakicked-1995,Billy-AndersonLocalization-2008,Roati-AndersonLocalization-2008,KondovAnderson-2011,Jotzu-Haldane-2014,ChiralCurrentsBloch-2014,Stuhl-Edge-2015,Fallani-chiral-2015,Ketterle-gauge-2015} simulators have made great progress in the experimental exploration of disordered and topological systems, while offering largely complementary capabilities and challenges. Photonic simulators generally permit control of system parameters and the detection of probability distributions at the microscopic, site-resolved level. However, the use of real materials as the medium for light transport makes these systems susceptible to inherent disorder in sample preparation~\cite{Hafezi-Robust-2014} and to absorption in the material~\cite{Absorption-2013}, and makes simulations in higher spatial dimensions and time-dependent control of system parameters non-trivial. For cold atoms, pristine and dynamically variable potential landscapes can be constructed based on their interaction with laser light. However, a microscopic control over system parameters is difficult to realize in atomic systems. Moreover, finite temperatures and the absence of hard-wall system boundaries have limited the observation of topological phenomena.

Here, we propose an atom optics-based~\cite{Adams-94,Berman-B97,Meystre-B01,Cronin-RMP} approach to the study of coherent transport phenomena, which incorporates many of the desired features of atomic and photonic experimental platforms. The scheme we describe is motivated in spirit by magnetic resonance-based techniques for local manipulation via global field addressing~\cite{MRI-1973,MeschedeMRI-2004}. In the context of studying transport phenomena, however, we consider an inhomogeneous landscape of site energies, with unique energy \emph{differences} between neighboring sites, which defines unique tunneling resonances for each site-to-site link. Combined with global field addressing that can drive transitions between neighboring sites, and in particular by simultaneous driving of many such transitions in an amplitude, frequency, and phase-controlled manner, this would allow for local control over the parameters of a discrete lattice model relevant to myriad coherent transport phenomena.

Atom optics offers a natural candidate system featuring a quadratic energy landscape and field-driven transitions between states. Here, we propose to create a discrete ``lattice of sites'' represented by free-particle momentum states of atomic matter waves, having quadratic energy-momentum dispersion, which can be effectively nearest-neighbor coupled via resonant two-photon Bragg transitions~\cite{Kozuma-Bragg,Denschlag-02}. The free particle dispersion allows for spectrally-resolved control over all parameters of the system at the single-link level, including all site-to-site ``tunneling'' amplitudes and phases, achieved by writing multiple radiofrequency sidebands onto a pair of interfering laser beams. We describe how this can enable the simulation of near-arbitrary single particle models, including two-dimensional Abelian U(1) lattice models describing integer Hall systems~\cite{vonKlitzing-IQHE-1986}. Additionally, we show how another well established atoms optics tool - stimulated Raman transitions that change both the internal state and momentum of atoms~\cite{Weiss-Raman-1993,Hagley-Raman-1999} - can be used to study non-Abelian U(2) gauge fields, which to date have been difficult to realize in photonic and cold atom settings.

The proposed scheme builds on a large body of work involving the study of transport phenomena using the evolution of momentum-space distributions of cold atomic gases~\cite{Moore-Qdeltakicked-1995,Hensinger-Phillips-DynTunnel,Steck-ChaosTunnel,Ryu-HighOrderRes-2006,Gadway-rotors}, including recent precision studies of the three-dimensional Anderson insulator-metal transition~\cite{Chabe-AndersonMetal-2008,Lemarie-CriticalStateAnderson-2010,Lopez-Universal}. While the majority of such studies have involved time-dependent driving by lattice potentials not fulfilling a resonant Bragg condition, notably in the realization~\cite{Moore-Qdeltakicked-1995,dArcy-QuantDiffusion-DKR-2001,Chabe-AndersonMetal-2008,Lemarie-CriticalStateAnderson-2010,Lopez-Universal} of quantum kicked rotor models~\cite{Fishman-Recurrences-1982,Altland-QKR-FieldTheory-1996}, here our proposed method operates deep within the resonant Bragg diffraction regime.

The paper is organized as follows. In Sec.~II, we introduce the basic experimental scheme based on state-preserving Bragg transitions that allows for the simulation of Abelian U(1) models in discrete lattice systems. In Sec.~III, we discuss in more depth some relevant aspects of the proposed scheme, including how it is extended to higher-dimensional systems, some of its unique capabilities, and some practical experimental limitations. In Sec.~IV, we introduce a second experimental scheme based on internal state-changing Raman transitions, which allows for the simulation of non-Abelian U(2) models. Finally, conclusions are presented in Sec.~V.

\section{II. ABELIAN U(1) LATTICE MODELS}

We begin by considering a generic system of two-level atoms, having a single internal ground (excited) state $| g \rangle$ ($| e \rangle$) with energy $\hbar \omega_{g(e)}$ and having a mass $M$. These two-level atoms and their interaction with a driving electric (laser) field $\mathbf{E}$, neglecting spontaneous emission, are described in the dipole approximation by the single particle Hamiltonian
\begin{equation}
\hat{H} = \frac{\hat{\mathbf{p}}^2}{2M} + \hbar \omega_e |e \rangle \langle e | + \hbar \omega_g |g \rangle \langle g | - \mathbf{d}\cdot \mathbf{E} \ ,
\label{EQ:e0}
\end{equation}
where $\mathbf{p}$ is the free particle momentum of the atoms and $\mathbf{d}=-|e|\mathbf{r}$ is the atomic dipole operator, with $\mathbf{r}$ a vector pointing from the atomic nucleus to the electron position. We assume that, as shown in Fig.~1(a), the electric field $\mathbf{E}$ of the driving lasers is composed of two distinct contributions -- a right-traveling field $\mathbf{E}^+(\mathbf{x},t)$ with a single frequency component and a left-traveling field $\mathbf{E}^-(\mathbf{x},t)$ with a number of discrete frequency components. Explicitly, we take these two fields to be
\begin{equation}
\mathbf{E}^+(\mathbf{x},t) = \mathbf{E}^+ \cos (\mathbf{k}^+ \cdot \mathbf{x} - \omega^+ t + \phi^+) \ \mathrm{and}
\label{EQ:e1a}
\end{equation}
\begin{equation}
\mathbf{E}^-(\mathbf{x},t) = \sum_{j} \mathbf{E}^{-}_{j} \cos (\mathbf{k}^{-}_{j} \cdot \mathbf{x} - \omega^{-}_{j} t + \phi^{-}_{j}) \ .
\label{EQ:e1b}
\end{equation}
We assume without loss of generality that the fields propagate along the x-axis, and moreover that they are nearly monochromatic such that $\mathbf{k}^+ = k \hat{x}$ and $\mathbf{k}^{-}_{j} \simeq -k \hat{x} \ \forall \ j$, with $k=2\pi/\lambda$ the wavevector of the laser light having wavelength $\lambda$. Similarly, all laser frequencies are detuned from atomic resonance ($\omega_{eg} \equiv \omega_e - \omega_g$) by a nearly equal amount $\Delta \equiv \omega_{eg} - \omega^+ \simeq \omega_{eg} - \omega^{-}_{j} \ \forall \ j$. For each frequency component of the driving electric field, we define the respective resonant Rabi couplings to be $\Omega^+ = - \langle e | \mathbf{d}\cdot \mathbf{E}^+ | g \rangle / \hbar$ and $\Omega^{-}_{j} = - \langle e | \mathbf{d}\cdot \mathbf{E}^{-}_{j} | g \rangle / \hbar$.

\begin{figure}[b]
\centering
\includegraphics[width=3.2in]{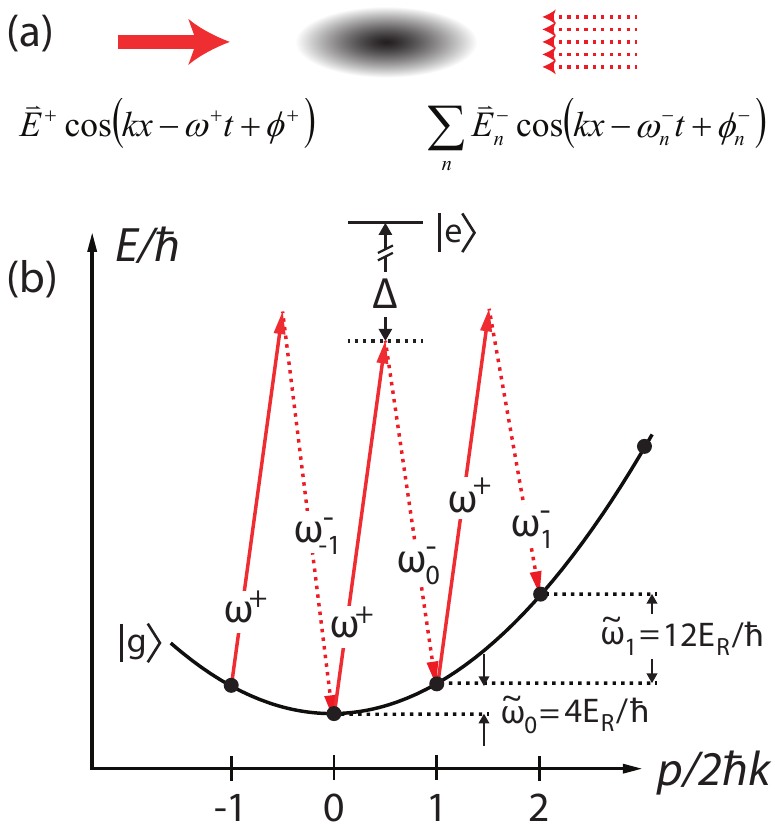}
\caption{(Color online) Experimental scheme for studying lattice-driven momentum-space dynamics.
(a)~Atomic matter waves are driven by a pair of counter-propagating laser fields, one of which is composed of several different frequency components, with controllable phase, frequency, and amplitude.
(b)~Energy-momentum dispersion. All laser fields are far-detuned by an amount $\Delta$ from atomic resonance between the ground $|g \rangle$ and excited $|e \rangle$ states. Stimulated two-photon Bragg transitions are driven by the pairs of interfering laser fields, coherently coupling plane-wave momentum states separated by two photon momenta ($2 \hbar k$). The quadratic free particle dispersion defines a unique two-photon Bragg resonance condition $\hbar\tilde{\omega}_n = (2n+1)4 E_R$ for each link between neighboring states. Each frequency component of the multi-frequency field addresses a unique state-to-state link.
}
    \label{FIG:fig1}
\end{figure}

Experimentally relevant terms related to the energy-momentum dispersion of the atoms are depicted in Fig.~1(b). The (nearly) common frequency detuning $\Delta$ of all the laser fields from atomic resonance is assumed to be much larger than all other relevant terms, including Doppler shifts of magnitude $|\mathbf{p}| k/M$ and the resonant Rabi coupling frequencies $|\Omega^+|$ and $|\Omega^{-}_j|$. This large single-photon detuning from resonance makes direct population of the atomic excited state $|e \rangle$ negligible. In the following we assume an effective ground-state Hamiltonian $\hat{H}_{\mathrm{eff}} = \hat{H}_0 + \hat{H}_{int}$ based on adiabatic elimination of the excited state $|e \rangle$. This effective Hamiltonian describes the free-particle kinetic energies $\hat{H}_0$ and light-atom interactions $\hat{H}_{int}$ that drive two-photon processes changing the atomic momenta by $\pm \hbar \mathbf{k}_{\mathrm{eff}} = \pm 2 \hbar k \hat{x}$ while leaving the internal state unchanged, characterized by virtual absorption of a photon from one laser field and stimulated emission into the other. Assuming that the atomic source is a condensate of atoms with small momentum spread $2\sigma_p \ll \hbar k$, we now define a discrete basis of relevant plane-wave momentum states $|n \rangle$ (with $n$ an integer), having momenta $\mathbf{p}_n = 2 n \hbar k \hat{x}$.

This discrete set of allowed momentum states will form the ``lattice of sites'' that can be coupled in a controlled way via two-photon transitions. These states have kinetic energies $E_n = \langle n | \hat{H}_0 | n \rangle = n^2 (4 E_R)$, where the single-photon recoil energy is given by $E_R = \hbar^2 k^2 / (2M)$. In the assumed form of the driving electric field $\mathbf{E}$, off-diagonal terms that increase the momentum by $2 \hbar k \hat{x}$ can in principle come about by absorption of a photon from the right-traveling field, followed by stimulated emission into any of the different frequency fields that constitute the left-traveling laser field. For such a process driven by the respective frequency component labeled by the index $j$, we define a corresponding two-photon Rabi coupling
\begin{equation}
\tilde{\Omega}_j e^{i \tilde{\phi}_j} = \frac{\Omega^{\ast -}_j \Omega^+}{2 \Delta} e^{i(\phi^+ - \phi^{-}_j)} \ ,
\label{EQ:e2}
\end{equation}
where $\tilde{\Omega}_j$ is assumed to be real and positive, and the phase shift associated with this process is determined by the phases $\phi^+$ and $\phi^{-}_j$ of the two laser fields, which can be easily controlled using acoust-optic or electro-optic modulators, for example.

We now define the effective ground-state Hamiltonian of this system in the interaction picture $\hat{H}_{\mathrm{eff}}^I$, where the time-dependence due to $\hat{H}_0$ is moved onto the system operators. In the ground-state plane-wave basis, the diagonal terms are now all zero (up to an ignored diagonal AC Stark shift that is common to all states). The nearest-neighbor off-diagonal elements, described in terms of the two-photon Rabi couplings for all allowed transitions, take the time-dependent form
\begin{equation}
\langle n + 1 | \hat{H}_{\mathrm{eff}}^I | n \rangle / \hbar = \sum_j \tilde{\Omega}_j e^{i \tilde{\phi}_j} e^{-i \delta^{(n)}_j t}  \ ,
\label{EQ:e3}
\end{equation}
where $\delta^{(n)}_j$ describes the two-photon detuning of the $j^{\mathrm{th}}$ frequency component from the $|n\rangle$ to $|n+1 \rangle$ transition, given as $\delta^{(n)}_j = (\omega^+ - \omega^{-}_j) - \tilde{\omega}_{n}$.
Here, the term $\tilde{\omega}_{n}$ describes the Doppler frequency shift of the transition $|n\rangle \rightarrow |n+1 \rangle$. Given that the free-particle dispersion is quadratic, its linear first derivative relates to a linearly varying Doppler frequency shift
\begin{equation}
\tilde{\omega}_{n} = \frac{\mathbf{p}_n \cdot \mathbf{k}_{\mathrm{eff}}}{M} + \frac{\hbar |\mathbf{k}_{\mathrm{eff}} |^2}{2 M} = (2n + 1) 4 E_R / \hbar  \ ,
\label{EQ:e3c}
\end{equation}
which serves to define the two-photon Bragg resonance condition for the $|n \rangle$ to $|n+1 \rangle$ transition.

We can make use of this unique state-to-state frequency shift to achieve the stated goal of controlling the off-diagonal elements in a link-specific manner. We explicitly assume that the two-photon detuning between each frequency component $j$ of the left-traveling field and the right-traveling field approximately satisfies a unique Bragg resonance condition. Formally, for every frequency component of the field labeled by index $j$, we set $\omega^+ - \omega^{-}_j \equiv \tilde{\omega}_j - \xi_j$, with $j$ an integer and $\xi_j$ a small ($\hbar \xi_j \ll 8 E_R \ \forall \ j$) and controllable detuning from the $j^\mathrm{th}$ two-photon Bragg resonance. This now brings us to the physical picture of building up individual links between a ``lattice'' of discrete momentum states, through the engineering of many interfering laser frequency components. In the limit of ``weak-driving'', which for this one-dimensional example we define as $\hbar \tilde{\Omega}_j \ll 8 E_R \ \forall \  j$, the bandwidth of two-photon transitions is sufficiently reduced such that at most one frequency component has a substantial contribution to each off-diagonal element. We then ignore all but the most near-resonant contribution for each off-diagonal coupling, in the spirit of a rotating wave approximation. This greatly simplifies the effective interaction-picture Hamiltonian, leading to weakly time-dependent off-diagonal couplings of the form
\begin{equation}
\langle n + 1 | \hat{H}_{\mathrm{eff}}^I | n \rangle / \hbar \approx \tilde{\Omega}_n e^{i \tilde{\phi}_n} e^{i \xi_n t}  \ .
\label{EQ:e4}
\end{equation}
For any two coupled modes, this weak time-dependence can be further absorbed into diagonal ``site''-energies $\varepsilon_n$ (related by $\xi_n = \varepsilon_{n+1} - \varepsilon_n$) by a rotating frame transformation, permitting a fully time-independent Hamiltonian description with a controlled ``potential landscape''. We will assume the less general case, however, where all frequency components of the applied fields \emph{exactly} fulfill a two-photon Bragg resonance condition, i.e. $\xi_n = 0 \ \forall \ n$. We then arrive at the desired description of a single particle tight-binding Hamiltonian
\begin{equation}
\hat{H}_{\mathrm{eff}}^I \approx \sum_n t_n (e^{i \varphi_n} \hat{c}^\dag_{n+1} \hat{c}_n + \mathrm{h.c.}) \ .
\label{EQ:e5}
\end{equation}
Here, arbitrary control over all tunneling amplitudes $t_n \equiv \hbar \tilde{\Omega}_n$ and tunneling phases $\varphi_n \equiv \tilde{\phi}_n$ of the system are enabled in a link-dependent way through control of a single global addressing field $\mathbf{E}^{-}(\mathbf{x},t)$. This can be simply accomplished, for example, by passing a single laser beam through a pair of acousto-optic modulators driven by tailored radiofrequency signals~\cite{Gemelke-IncMottDomains-2009}. Moreover, the tailored radiofrequency signal can be smoothly varied in time, such that the parameters of the model system can be made time-dependent.

\begin{figure}[t]
\centering
\includegraphics[width=3.3in]{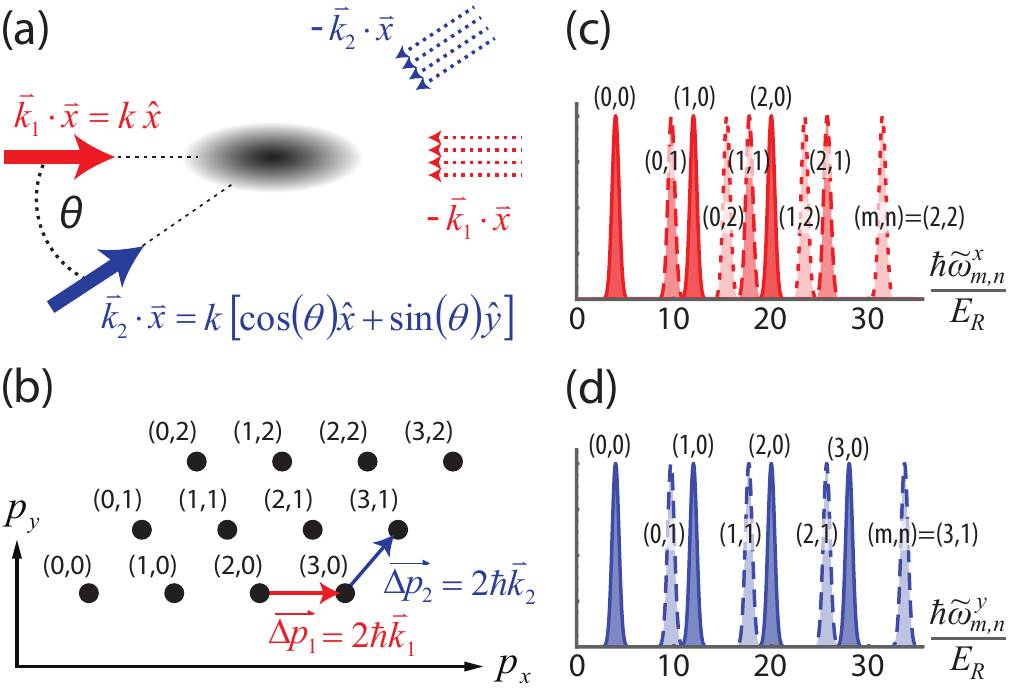}
\caption{(Color online) A two-dimensional lattice system with spectrally resolved link resonances.
(a)~Two pairs of interfering laser fields (set 1 shown in red, set 2 shown in blue), intersecting in a plane at an angle $\theta$, are shone onto a collection of atomic matter waves. Cross interferences between the two pairs of beams can be avoided by choice of laser polarizations or by introducing a frequency offset between the two pairs.
(b)~The discrete ``lattice'' of momentum states that can be populated by stimulated two-photon Bragg transitions, starting from zero momentum.
(c)~The spectral positions of the nearest-neighbor Bragg resonances $\tilde{\omega}^x_{m,n}$ driven by the laser pair 1, relating to the finite-sized set of states labeled $(m,n)$ with momenta $\mathbf{p}_{m,n} = 2\hbar (m \mathbf{k}_1 + n \mathbf{k}_2)$ as shown in (b).
(d)~Same as in (c), but for the Bragg transition resonances $\tilde{\omega}^y_{m,n}$ addressed by the second pair of lasers.
}
    \label{FIG:fig2}
\end{figure}

The scheme as described, with local control over tunneling amplitudes and phases, permits the study of near-arbitrary one-dimensional systems. Of natural interest would be the study of superlattice systems known to have non-trivial topological properties~\cite{Jackiw-Rebbi-1976,SSH-1979,Quasi1D-2D-Chen-2012,Quasi1D-2D-Kraus-2012A,Quasi1D-2D-Kraus-2012B}, in particular when combined with either additional modulation of the tunneling parameters~\cite{Ganeshan-ZEM-2013,Ganeshan-Weyl-2015,Chen-Kaleid-2015} or in the presence of disorder~\cite{TaylorHughes-ChiralAIII,Prodan-AIIIBDI-2014}.

\section{III. FURTHER ASPECTS OF THE SCHEME}

\subsection{A. Extension to higher dimensions}

While the ability to simulate arbitrary Hamiltonians describing lattice transport in one dimension would allow for a number of interesting studies, particularly relevant to disordered and symmetry-protected topological states, the tunneling phases $\varphi_n$ are of little physical consequence when applied only to one-dimensional systems with nearest-neighbor couplings. In higher dimensions, a natural application of the ability to engineer link-specific phases would be to mimic the Aharonov-Bohm phase $\phi_{AB}$ acquired by charged particles (with charge $q$) moving along a path $P$ in an electromagnetic vector potential $\vec{A}$, $\phi_{AB} = (q/\hbar) \int_P \vec{A} \cdot \vec {dx}$. This would allow the study of topologically non-trivial (2+1)-dimensional Abelian U(1) models, such as those describing the integer quantum Hall effect exhibited by electrons confined in two dimensions under the influence of strong transverse magnetic fields~\cite{vonKlitzing-IQHE-1986,ChalkCodd-IQHE}. The local manipulation of phases could also allow the study of random magnetic flux models~\cite{Lee-Fisher-RandomFlux-1981,Altland-RandomFlux}, which are believed to exhibit metallic behavior and provide an interesting counterexample to Anderson's theorem~\cite{Anderson-DiffRand-1958,Abrahams-Scaling-1979} in two dimensions. Higher-dimensional studies allow access to novel lattice geometries as well, where link-specific control over tunneling amplitudes can be used to transform a simple square lattice into a brick-wall honeycomb lattice~\cite{Tarruell-Dirac-2012} by setting certain links to zero tunneling. In general this control allows one to impose hard-wall boundary conditions, and a two-dimensional scheme with tailored links would allow one to create one-dimensional systems with periodic boundary conditions. Recently, researchers have used such a local manipulation in photonic simulators to probe novel questions about the bulk-boundary correspondence in integer quantum Hall systems~\cite{Hafezi-GaugedEdges-2015}.

Here, we describe the simple extension to realizing two dimensional models that preserve full spectral control over all tunneling links (with straightforward extensions to higher-dimensional systems as well). We consider the case of driving by two independent pairs of counter-propagating laser fields as shown in Fig.~2, where we neglect any effect of cross interferences (by choice of polarization or an appropriate frequency offset). Elementary changes to the atomic momentum by $\mathbf{\Delta p}_1 = 2\hbar \mathbf{k}_1$ and $\mathbf{\Delta p}_2 = 2 \hbar \mathbf{k}_2$ result from allowed two-photon Bragg processes as in the earlier-described scheme. Assuming that we start with population nominally at zero momentum, this defines a set of possible momentum states $|m,n \rangle$, having momenta $\mathbf{p}_{m,n} = 2\hbar (m \mathbf{k}_1 + n \mathbf{k}_2)$. We next assume, without loss of generality, that $\mathbf{k}_1 = k_1 \hat{x}$ and $\mathbf{k}_2 = k_2 [\cos \theta\hat{x} + \sin \theta \hat{y}]$ as depicted in Fig.~2. If $k_1 \neq k_2$, this can allow for effectively higher-dimensional systems (of finite extent) to be realized even for $\theta = 0$~\cite{Gadway-rotors}. Here we consider instead the case of driving by lattices with near-identical wavevectors along two different directions, i.e. $k_1 \simeq k_2 = k$ and $\theta \neq 0,\pi$. The resulting kinetic energies of the $|m,n \rangle$ states will be given by
\begin{equation}
E_{m,n} = 4 E_R [m^2 + n^2 + 2mn \cos\theta] \ .
\end{equation}
So long as the lattice directions are not orthogonal ($\theta \neq \pi/2 , 3\pi/2$), there will exist unique Bragg resonance conditions for each link of a finite-sized two-dimensional system.

\begin{figure*}
\centering
\includegraphics[width=6.4in]{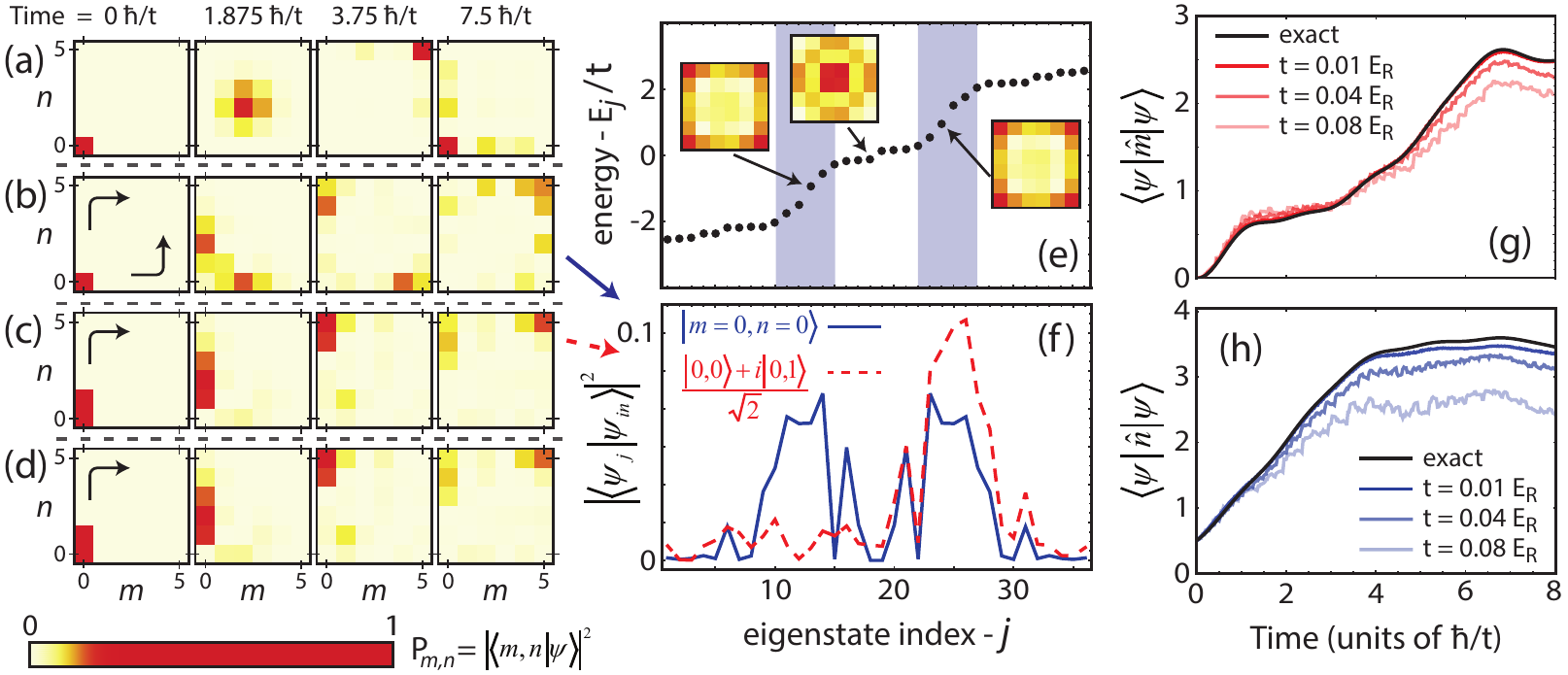}
\caption{(Color online) Simulated momentum-space dynamics of a small (6 site $\times$ 6 site) two-dimensional (2D) lattice system.
(a-d)~Probability distributions of the different momentum modes $|m,n \rangle$, following dynamics initiated from a state $|\psi_{in} \rangle$, at times of 0, 1.875, 3.75, and 7.5 in units $\hbar / t$, with $t$ the tunneling energy.
(a)~Shown for a regular lattice with homogeneous tunneling energies $t$ and no tunneling phases, starting from $|\psi_{in} \rangle = |0,0\rangle$. The dynamics shown relate to evolution governed by Eq.~(11) from the text.
(b)~Same, but for an enclosed synthetic magnetic flux of $2\pi/3$ per lattice plaquette, set through a non-trivial tunneling phase along one direction, $\varphi^y_{m,n} = 2 m \pi / 3$. In this case, the particles avoid entering the bulk or interior of the system, and instead propagate along the system boundaries. Because state preparation is based on mode projection, with no explicit energy dependence, a combination of clockwise and counter-clockwise propagating edge states are populated.
(c)~As in (b), but starting from the state $|\psi_{in} \rangle = (|0,0\rangle + i |0,1\rangle)/\sqrt{2}$. The populated state propagates with essentially only one chirality.
(d)~Exactly as in (c), but including all tunneling contributions due to the entire sideband spectrum [i.e. with dynamics governed by the 2D equivalent of Eq.~(5) and not Eq.~(11), with $t/E_R = 0.01$].
(e)~Energy spectrum relating to the systems of (b) and (c), with $2\pi/3$ flux enclosed per lattice plaquette. The system is split into 3 bulk energy bands, and features additional dispersive edge states (shaded in blue as a guide to the eye). Insets show the modal distribution of different energy eigenstates.
(f)~Probability distribution of eigenstates populated by projection from $|\psi_{in} \rangle = (|0,0\rangle$ (solid blue) and $|\psi_{in} \rangle = (|0,0\rangle + i |0,1\rangle)/\sqrt{2}$ (dashed red).
(g,h)~Center-of-mass position dynamics, in terms of mode numbers $m$ and $n$ along the two directions, for enclosed flux and initial state as in (c) and (d). The black lines show the exact dynamics as in (c). From darker to lighter colors, the red [blue] lines in (g) [(h)] show dynamics for $t/E_R = 0.01$, 0.04, 0.08. The smallest energy gap between spectral resonances $\tilde{\omega}^{x(y)}_{m,n}$ in the system is $0.97 E_R$.
}
    \label{FIG:fig3}
\end{figure*}

Similar to the unique Bragg transition frequencies $\tilde{\omega}_n$ between adjacent states $|n \rangle$ and $|n+1 \rangle$ in one dimension, described in Eq.~\ref{EQ:e3c}, in two dimensions we have unique Bragg transition frequencies that depend on the initial state $|m,n \rangle$ and in which direction the momentum is imparted. For a momentum change of $\mathbf{\Delta p}_1$, this gives the condition
\begin{equation}
\tilde{\omega}^x_{m,n} = [2m + 1 + 2 n \cos \theta] 4 E_R / \hbar  \ .
\label{EQ:6a}
\end{equation}
A similar condition ($\tilde{\omega}^y_{m,n} = [2n + 1 + 2 m \cos \theta] 4 E_R / \hbar$) exists for a momentum change $\mathbf{\Delta p}_2$,
and because there is no cross interference between the pairs of laser fields, unique spectral control of tunneling terms along all links can still be preserved even if there exist overlapping resonances $\tilde{\omega}^x_{m,n} = \tilde{\omega}^y_{m',n'}$ along the two different directions. Following the procedure as in Sec.~II, through the application of spectral sidebands to one laser from each pair, being controlled in amplitude and phase and offset in frequency from the counter-propagating partner to fulfill particular resonance conditions, one may realize a two-dimensional Abelian U(1) lattice model of the form
\begin{equation}
\begin{split}
\hat{H}_{\mathrm{eff}}^I \approx \sum_{m,n} [ t^x_{m,n} (e^{i \varphi^x_{m,n}} \hat{c}^\dag_{m+1,n} \hat{c}_{m,n} + \mathrm{h.c.}) \\
+ \ t^y_{m,n} (e^{i \varphi^y_{m,n}} \hat{c}^\dag_{m,n+1} \hat{c}_{m,n} + \mathrm{h.c.}) ]  \ .
\end{split}
\end{equation}

As a concrete example, we analyze in Fig.~3 the effective dynamics that can be driven in a two-dimensional system with non-trivial tunneling phases, relating to an effective Aharonov--Bohm phase acquired by particles evolving in the system of momentum states. We show that far in the weak-driving limit, the effective dynamics that emerge from Eq.~(5) exactly coincide with those of Eq.~(11). In the case of a non-zero synthetic magnetic flux, these dynamics show insulating behavior in the bulk of the system and transport along the edge of the system. The dynamics illustrated in Fig.~3 also highlight an important aspect of the simplest studies that can be performed using the proposed scheme - those involving population initiated in one or a few momentum states, with laser-driving turned on suddenly. Similar to the case of many photonic simulators~\cite{SzameitReview-2010}, spatial projection onto the system's eigenmodes dictates the ensuing dynamics, and there is no explicit energy selection or preparation in the system's ground state. For a non-zero synthetic magnetic flux, population initiated in the bulk of the system will remain stationary, while population on the system’s edge will undergo transport. Furthermore, Fig. 3 shows how particular edge modes can be populated by beginning with population in a superposition of multiple momentum states. Given the similarities to photonic systems, with respect to projective state initialization and out-of-equilibrium dynamics, we expect that many of the techniques developed for studying topological properties of photonic simulators should prove useful in the envisioned atom optics setting~\cite{DetectionTechniques-TopTransition-Silberberg-2013}.

One issue to note in accessing higher-dimensional models is that the frequency spacing between the link-specific Bragg resonances, found in one dimension to have the value $8 E_R / \hbar$, is reduced as the number of links in each direction is increased. This in general requires lower tunneling rates (two-photon Rabi rates) to remain in the weak-driving limit where individual resonances are spectrally resolved. Practically, a more realistic approach to studying higher dimensional systems while preserving arbitrary control of all parameters may be found in systems extended in one direction and with only two or a few sites along a second~\cite{Chiral-Paredes,ChiralCurrentsBloch-2014,Stuhl-Edge-2015,Fallani-chiral-2015} or second and third~\cite{Loss-Frac-Ferm-2013} direction.

\subsection{B. Unique features}

The suggested atom optics-based approach, which allows for precise and time-dependent control of a single particle lattice model at a link-specific level, affords many unique experimental capabilities relevant to quantum simulation. Furthermore, the fact that the effective ``tunneling'' transitions between sites are explicitly field-driven and do not result from quantum tunneling through a barrier allows in principle for several unique features. As discussed in the previous section, this allows for the simulation of higher-dimensional systems of finite extent in three or fewer physical dimensions. It additionally allows for direct and independent control of tunneling terms beyond nearest-neighbor. For example, one can access next-nearest-neighbor hopping terms by driving second-order Bragg processes with resonances given by $2\tilde{\omega}^{(NNN)}_{n} = (4n +4)4 E_R$, which are spectrally distinct from the first-order resonances~\cite{Kozuma-Bragg}. Controlled access to such terms would allow tunable symmetry breaking (inversion or particle-hole) of topological insulator systems. Additionally, it has been shown~\cite{NNN-ShuChen-2014} that the combination of nearest-neighbor (NN) and next-nearest-neighbor (NNN) tunneling in one dimension can be used to realize systems analogous to the two-dimensional Haldane model~\cite{Haldane-HaldaneModel-1988}, allowing study of the anomalous quantum Hall effect in an experimentally simple setting. Such a combination of terms may also allow for the study of Lifshitz-type behavior~\cite{LeHur-Lifshitz-2013}, e.g. as found in axial next-nearest-neighbor Ising (ANNNI) models~\cite{Elliot-ANNNI-1961,Fisher-Selke-ANNNI-1980,Selke-ANNNI-Review-1988}.

Another relatively unique aspect of the proposed system stems from the combination of local and time-dependent parameter control. To note, either local control or time-dependent control of the system parameters would allow, e.g., for the controlled implementation of quenched disorder or of a time-varying Hamiltonian for quantum annealing to novel ground states~\cite{Multifractal-Aoki-1983,EversMirlinReview-2008}, respectively. In this context, their specific combination could allow for the study of \emph{annealed disorder}, with randomly distributed system parameters that are additionally modulated in time~\cite{Osterloh-NonAbel-2005PRL}. In essence, such modulation of the lattice parameters over an appropriate range of frequencies can mimic the coupling of particles to a thermal phonon bath. Through the modulation of disorder at frequencies corresponding to relevant energy scales of the model system being studied, such annealed disorder could allow access to the thermodynamic properties of an otherwise intrinsically out-of-equilibrium system.

\subsection{C. Limitations}

There exist several practical limitations to the timescales over which the proposed scheme can be used to simulate coherent dynamics. The major limitation comes from the fact that ultracold atoms are not idealized zero-momentum plane waves, but have a spread in momentum due to finite temperature, interactions between particles, and the zero-point motion associated with the ground state of their confining potential~\cite{Stenger-Bragg}. The momentum spread of trapped Bose--Einstein condensates is typically much smaller than the recoil momentum, $2\sigma_p \ll \hbar k$, such that the picture of a discrete lattice of states is justified. However, even a small but finite momentum spread will introduce restrictions on the experimental timescales over which coherent momentum-space dynamics can be observed. Coherent dynamics in momentum-space requires that momentum states with direct off-diagonal coupling occupy indistinguishable spatial modes. In other words, the laser-driven dynamics will occur only in the near-field regime~\cite{Deng-Talbot-1999,Ryu-HighOrderRes-2006}, before the populated momentum states have time to spatially separate into distinct wavepackets.

While this imposes a strong limit on the timescales over which coherent transport phenomena can be expected to occur, a significant number of coherent tunneling events can still be achieved. Moreover, this effect will be less relevant to the observation of phenomena involving localized states or ballistic, non-dispersive propagation in momentum-space. Still, we can provide a lower estimate for the limiting timescale based on the worst case scenario, the Ramsey decoherence time in the absence of continuous coupling between two states. For nearest-neighbor states differing in velocity by $2 v_R$ (with $v_R \equiv \hbar k / M$ the recoil velocity), their spatial overlap will be lost roughly on the timescale $T_{coh} = L_c / 2 v_R$, where $L_c$ is the cloud's spatial coherence length along the direction of momentum transfer. We assume that $L_c$ is determined at ultracold temperatures and low densities by the finite system size in a trapping potential, and we relate this to the number of lattice sites $N_s$ (of the interfering laser fields) over which the atomic distribution would extend, with $L_c = N_s \lambda / 2$. This description allows for the simple relation $T_{coh} = N_s \tau_{0}$, where $\tau_{0} = h / 8 E_R$. We recall that in one dimension the tunneling rates are restricted to be much less than $8 E_R / \hbar$ to spectrally resolve individual link resonances. Assuming tunneling rates $t \sim 8 E_R / 10 \hbar$, we can expect Ramsey coherence times corresponding to roughly $N_s / 10$ tunneling events.

This expected limitation to the scheme motivates some practical considerations. When implementing higher-dimensional ``lattices'' of momentum states, because the tunneling rates necessary to achieve complete spectral control of all tunneling parameters become severely restricted, the dynamics will remain coherent for far fewer tunneling events. We thus expect that it will be more realistic to pursue studies of one-dimensional lattice and superlattice systems, as well as ladder-type systems with only a few sites along a second direction~\cite{Chiral-Paredes,ChiralCurrentsBloch-2014,Stuhl-Edge-2015,Fallani-chiral-2015} or two additional directions~\cite{Loss-Frac-Ferm-2013}. Additionally, an active increase of the relevant experimental timescales may be achieved by increasing the spatial coherence length of the atomic sample prior to the lattice-driven dynamics. This can be achieved by an adiabatic decrease of the trapping depth and stiffness, leading to an increase of the atoms' spatial extent~\cite{Kozuma-Bragg,Leanhardt12092003}. One can also pursue still more active methods for increasing of the atomic sample's size based on analogies to Gaussian beam optics, namely by using matter-wave lensing techniques~\cite{Chu-DeltaKick} for the construction of an atomic beam expander or Galilean telescope. Such techniques have recently been employed to create mm-scale atomic clouds of $^{87}$Rb with pK-scale temperatures~\cite{Kasevich-Lensing-2014}, which for lattice light tuned near the D$_2$ transition would allow for a few hundred coherent tunneling events in one dimension.

If these studies are performed in atomic free fall or free expansion, so as to minimize any influence of trapping potentials on the ensuing matter-wave dynamics, another practical limitation is found. Assuming a geometry of lattice-driving along a direction perpendicular to gravitational acceleration, to avoid additional complications due to time-varying Doppler shifts, then gravity will cause the atoms to fall away from the region of light-atom interaction. Restricting the atoms to fall less than $d_0=1$~mm, for example, will restrict the experimental timescales to $T_{grav} = \sqrt{2 d_0 / g} \sim 14$~ms (where $g = 9.81 \ \mathrm{m}/\mathrm{s}^2$ is the assumed gravitational acceleration due to free fall), or roughly 270 tunneling events in the case of one-dimensional simulations. These timescales are generally less restrictive than those due to the near-field constraint, and can be largely assuaged through levitation in a magnetic field gradient without introducing significant external confinement.

Lastly, we remark that \emph{spontaneous} photon scattering can in principle provide an additional limitation to the observation of coherent momentum-space dynamics driven by \emph{stimulated} photon scattering~\cite{Grimm-dipole-2000}. Practically, however, the heating rates due to off-resonant absorption and re-emission events can be mitigated by setting the single-photon detuning $\Delta$ to be large compared to the spontaneous decay rate $\Gamma$ of the excited state $|e \rangle$.

\section{IV. NON-ABELIAN U(2) LATTICE MODELS}

We now describe a straightforward extension to the scheme described in Sec.~II, which is based on using internal state-changing two-photon Raman transitions~\cite{Weiss-Raman-1993,Hagley-Raman-1999} as opposed to state-preserving Bragg transitions. This modified scheme requires the use of two low energy internal ground states $|g_1\rangle$ and $|g_2 \rangle$, such as two $|m_F = 0 \rangle$ Zeeman sublevels of different hyperfine manifolds, as typically used in Raman atom interferometers~\cite{Berman-B97,Weiss-Raman-1993}. At low magnetic field, these states have an energy difference $\hbar \omega_{12} \equiv \hbar \omega_{g_2} - \hbar \omega_{g_1}$ determined by their hyperfine splitting, which we assume greatly exceeds the largest kinetic energy scales in the problem. As we show below, this extra internal ground state degree of freedom, when combined with a laser-driving protocol similar to that described in Sec.~II, will allow for the study of U(2) lattice models with near-arbitrary parameter control.

\begin{figure}[b]
\centering
\includegraphics[width=3.2in]{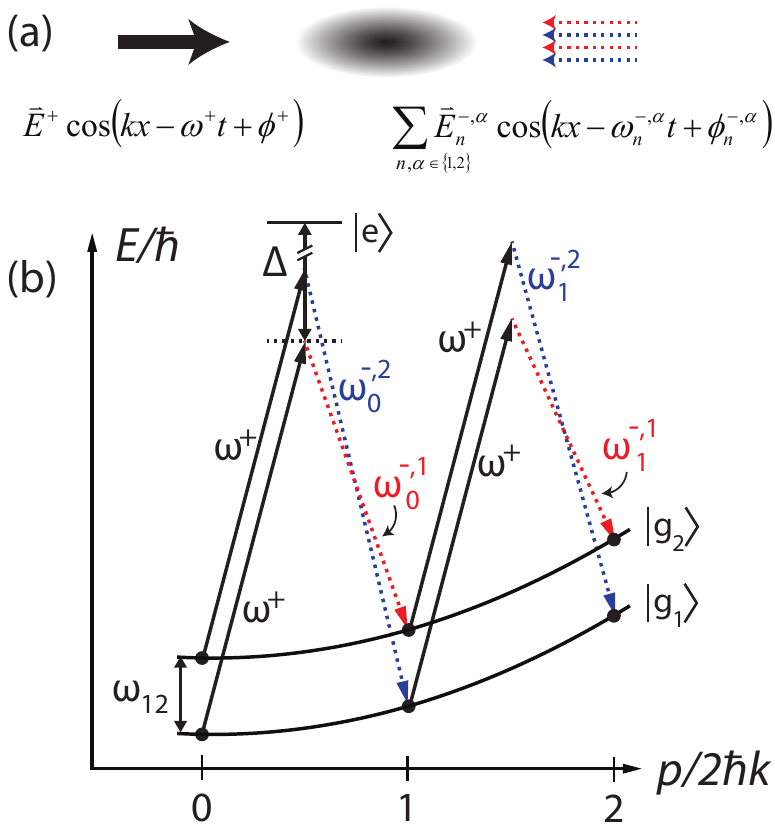}
\caption{(Color online) Laser driving scheme for studying U(2) lattice dynamics.
(a)~Counter-propagating laser fields drive a sample of atomic matter waves, where the field along one direction is composed of multiple spectral components, having frequencies $\omega_n^{-,\alpha}$.
(b)~Energy-momentum dispersion diagram. Two low-energy internal states $|g_1 \rangle$ and $|g_2 \rangle$ are coupled through stimulated state- and momentum-changing two-photon Raman transitions. All laser fields are far-detuned by an amount $\Delta \gg \omega_{12}$ from atomic resonance, so that the excited state $|e\rangle$ is only virtually driven. For the left-traveling, multi-frequency field, two distinct sets of frequency components (labeled $\alpha = 1$ and 2), are used in conjunction with the right-traveling field to drive unique state- and momentum-changing transitions that depend on the initial internal state, as described in the text and shown in the figure.
}
    \label{FIG:fig4}
\end{figure}

We consider interaction of these three-level atoms (having mass $M$) with an electric field $\mathbf{E}$, governed by
\begin{equation}
\hat{H} = \frac{\hat{\mathbf{p}}^2}{2M} + \hbar \omega_e |e \rangle \langle e | + \sum_{\alpha \in \{1,2\} } \hbar \omega_{g_\alpha} |g_\alpha \rangle \langle g_\alpha | - \mathbf{d}\cdot \mathbf{E} \ .
\label{EQ:e0b}
\end{equation}
We assume from the outset that the electric field is formed by two laser fields counter-propagating along the x-axis, a right-traveling field $\mathbf{E}^+(\mathbf{x},t)$ and a left-traveling field $\mathbf{E}^-(\mathbf{x},t)$, having nearly identical wavevector magnitudes $k$. As in the previous scheme, the right-traveling field is monochromatic ($\omega^+$) and far-detuned from atomic resonance by an amount $\Delta \equiv (\omega_e - \omega_{g_1}) - \omega^+ \gg \omega_{12}$. The left-traveling field contains a number of spectral components with frequencies $\omega^{-,\alpha}_n$. The fields are explicitly given by
\begin{equation}
\mathbf{E}^+(\mathbf{x},t) = \mathbf{E}^+ \cos (kx - \omega^+ t + \phi^+) \ \mathrm{and}
\end{equation}
\begin{equation}
\mathbf{E}^-(\mathbf{x},t) = \sum_{n,\alpha \in \{1,2\}} \mathbf{E}^{-,\alpha}_{n} \cos (-kx - \omega^{-,\alpha}_{n} t + \phi^{-,\alpha}_{n}) \ .
\end{equation}
As before, the index $n$ will relate to transitions between plane-wave states with momenta $2n\hbar k$ and $2 (n+1)\hbar k$. The index $\alpha = 1$ relates to processes where atoms undergo a transition from $|g_1\rangle$ to $|g_2\rangle$ as their momentum increases by $2\hbar k$ ($|g_1 , n \rangle \leftrightarrow |g_2 , n+1 \rangle$), while $\alpha = 2$ relates to momentum-increasing processes that transition from $|g_2\rangle$ to $|g_1\rangle$ ($|g_2 , n \rangle \leftrightarrow |g_1 , n+1 \rangle$), as depicted in Fig.~4. Making the restrictive assumption that every frequency component is \emph{exactly} resonant with a unique momentum-changing Raman transition, the frequencies of the left-traveling field's various components are given by
\begin{equation}
\omega_n^{-,1} = \omega^+ - \omega_{12} - (2n+1)4 E_R / \hbar \ \ \mathrm{and}
\end{equation}
\begin{equation}
\omega_n^{-,2} = \omega^+ +\omega_{12} - (2n+1)4 E_R / \hbar \ \ .
\end{equation}
The relevant one-photon Rabi frequencies relating to interaction with the different field components are given by $\Omega^{+,\alpha} = - \langle e | \mathbf{d}\cdot \mathbf{E}^+ | g_\alpha \rangle / \hbar$, $\Omega^{-,1}_{n} = - \langle e | \mathbf{d}\cdot \mathbf{E}^{-,1}_{n} | g_2 \rangle / \hbar$, and $\Omega^{-,2}_{n} = - \langle e | \mathbf{d}\cdot \mathbf{E}^{-,2}_{n} | g_1 \rangle / \hbar$. As in the previous case, we assume that we are in the limit where all one-photon Rabi frequencies are much less than the single-photon detuning $\Delta$. This restriction allows us to again consider an adiabatic elimination of the excited state $|e\rangle$, with only stimulated two-photon processes allowed. For processes characterized by absorption of a photon from the right-traveling laser field and stimulated emission into the frequency component of the left-traveling field with indices $n$ and $\alpha$, the effective two-photon Rabi frequency and phase shift are given by
\begin{equation}
\tilde{\Omega}^\alpha_n e^{i \tilde{\phi}^\alpha_n} = \frac{\Omega^{\ast -,\alpha}_n \Omega^+}{2 \Delta} e^{i(\phi^+ - \phi^{-,\alpha}_n)} \ .
\end{equation}

We again make the stronger restriction that the two-photon Rabi frequencies are all smaller in magnitude than the frequency spacing between unique spectral components, $\hbar \tilde{\Omega}^\alpha_n \ll 8 E_R / \hbar \ \forall \ n , \alpha$. In this weak-driving limit, the off-diagonal elements of the interaction Hamiltonian $\hat{H}_{\mathrm{eff}}^I$ have only one dominant contribution
\begin{equation}
\langle g_2, n + 1 | \hat{H}_{\mathrm{eff}}^I | g_1, n \rangle / \hbar \approx \tilde{\Omega}^1_n e^{i \tilde{\phi}^1_n}  \ \ \mathrm{and}
\end{equation}
\begin{equation}
\langle g_1, n + 1 | \hat{H}_{\mathrm{eff}}^I | g_2, n \rangle / \hbar \approx \tilde{\Omega}^2_n e^{i \tilde{\phi}^2_n}  \ \ .
\end{equation}
The dynamics of this system, neglecting differential AC Stark shifts of the two ground states, can again be described by an effective tight-binding Hamiltonian in the limit of weak-driving, given by
\begin{equation}
\begin{split}
\hat{H}_{\mathrm{eff}}^I \approx \sum_n t^{+}_n (e^{i \varphi^{+}_n} \hat{c}^\dag_{n+1} \hat{\sigma}_+ \hat{c}_n + \mathrm{h.c.}) \\
+ \sum_n t^{-}_n (e^{i \varphi^{-}_n} \hat{c}^\dag_{n+1} \hat{\sigma}_- \hat{c}_n + \mathrm{h.c.}) \ \ ,
\end{split}
\end{equation}
with $\hat{\sigma}_+ = (\hat{\sigma}_x + i \hat{\sigma}_y) / 2 = |g_2\rangle \langle g_1 |$ and $\hat{\sigma}_- = (\hat{\sigma}_x - i \hat{\sigma}_y) / 2 = |g_1\rangle \langle g_2 |$, where $\hat{\sigma}_{x}$ and $\hat{\sigma}_{y}$ are the Pauli matrices. Control of the laser sideband amplitudes and phases provides arbitrary control over all tunneling amplitudes $t^{+}_n \equiv \hbar\tilde{\Omega}^1_n$ and $t^{-}_n \equiv \hbar\tilde{\Omega}^2_n$ and tunneling phases $\varphi^{+}_n \equiv \tilde{\phi}^1_n$ and $\varphi^{-}_n \equiv \tilde{\phi}^2_n$. For every site-to-site transition, there exist two possible pathways involving non-commuting operations on the internal (pseudo)spin degree of freedom. By coordination of the tunneling amplitudes and phases relating to each of these pathways, a tunable U(2) lattice model can be constructed. To be explicit, if we assume equal tunneling amplitudes for the two pathways ($t^+_n = t^-_n \equiv t_n$), the effective Hamiltonian can be recast as
\begin{equation}
\hat{H}_{\mathrm{eff}}^I \approx \sum_n t_n (\hat{c}^\dag_{n+1} \hat{U}_n \hat{c}_n + \mathrm{h.c.}) \ \ ,
\end{equation}
where $\hat{U}_n = e^{i\Phi_n / 2} [\cos(\Theta_n / 2)\hat{\sigma}_x - \sin(\Theta_n / 2)\hat{\sigma}_y ]$, with $\Phi_n = \varphi^+_n +\varphi^-_n$ and $\Theta_n = \varphi^+_n -\varphi^-_n$. This allows us to vary the U(1) phase and SU(2) internal state spin-rotation associated with every individual tunneling link. Further inclusion of state-preserving Bragg transitions associated with each link would allow for an even more generalized form of the $\hat{U}_n$ matrices.

Following the procedure outlined earlier, this U(2) lattice model can also be performed in more than one spatial dimension, allowing for a model of the form
\begin{equation}
\begin{split}
\hat{H}_{\mathrm{eff}}^I \approx \sum_{m,n} [ t^x_{m,n} (\hat{c}^\dag_{m+1,n} \hat{U}^{x}_{m,n} \hat{c}_{m,n} + \mathrm{h.c.}) \\
+ t^y_{m,n} (\hat{c}^\dag_{m,n+1} \hat{U}^{y}_{m,n} \hat{c}_{m,n} + \mathrm{h.c.}) ]  \ \ .
\end{split}
\end{equation}
This allows for the study of genuine non-Abelian U(2) models, where motion along closed paths can lead to non-trivial operations on the atoms' internal degree of freedom. For the smallest counter-clockwise path around a four site plaquette, this can lead to an operation distinct from identity $I$,
\begin{equation}
\hat{U}^{\circlearrowleft}_{m,n} \equiv \hat{U}^{\dagger y}_{m,n}\hat{U}^{\dagger x}_{m,n+1}\hat{U}^{y}_{m+1,n}\hat{U}^{x}_{m,n} \neq e^{i\beta} I \ \ ,
\end{equation}
such that the Wilson loop variable associated with this closed path, tr($\hat{U}^{\circlearrowleft}_{m,n}$), is not equal to 2, the dimension of the internal state space. Independent control over all tunneling amplitudes and phases allows for the study of models with homogeneous Wilson loops~\cite{ConstantWilsonLoop-2009} for all elementary lattice plaquettes, as well as spatially varying and disordered configurations. In particular, it has been suggested that the U(2) random flux model may be of direct relevance to the effect of giant magnetoresistance displayed in manganese oxides~\cite{Altland-RandomFlux}. Furthermore, while the described setup is clearly restricted to the simulation of matter interacting with \emph{classical} Abelian and non-Abelian gauge fields, Ref.~\cite{Osterloh-NonAbel-2005PRL} recently raised the interesting prospect of using such simulations - along with averaging over an appropriate distribution of static and annealed classical gauge field configurations - to gain insight into certain properties of lattice gauge theories describing the interaction of matter with \emph{dynamical} gauge fields.

\section{V. CONCLUSIONS}

In conclusion, we have presented a simple experimental scheme for studying nearly arbitrary single particle transport phenomena based on well established atom optics techniques. We described two variations of this scheme, based on internal state-preserving Bragg transitions and internal state-changing Raman transitions, which enable the study of Abelian U(1) and non-Abelian U(2) lattice models, respectively. Some unique features of this platform were discussed, including the possibilities of studying annealed disorder and variable-range hopping. We have discussed practical limitations to the timescales of coherent evolution that this scheme allows, which relate to several tens to several hundreds of tunneling events for realistic system parameters. We neglected discussion of further extensions, such as the use of additional internal ground states for the simulation of U($N$) models with $N>2$, and we neglected a discussion of the important and intriguing role of nonlinear interactions between the atoms themselves.

In contrast to many photonic simulators, a system of atomic condensates evolving in momentum space would naturally play host to significant nonlinear processes~\cite{Rolston-NL-2002}, such as cross-phase modulation, self-phase modulation, and four-wave mixing~\cite{Pertot-10-PRL}. Moreover, the general scheme of developing link-resolved control of tunneling by use of an inhomogeneous potential and global field addressing may be transportable to strongly-correlated studies. In an optical lattice simulator, for example, tunable inhomogeneous potentials may be created by projective methods~\cite{Salomon-projective-1998,Bakr-QGM-2009,Chin-Roton-2015}, and global addressing via laser-assisted tunneling~\cite{Sias-PhotonAssisted-2008,Ma-PhotonAssisted-2011} may be used to reintroduce site-to-site coupling in a link-dependent fashion, allowing local control over tunneling amplitudes and phases.

Lastly, as a natural consequence of developing a new atom optics-based system for simulating coherent transport phenomena, the atom optics toolset will be expanded to include unique new capabilities for the manipulation of atomic matter waves.

\section{ACKNOWLEDGMENTS}

We thank Taylor Hughes for helpful conversations, and Brian DeMarco for helpful conversations and comments.

%

\end{document}